\documentclass[prd,aps,a4paper,superscriptaddress,twocolumn,nofootinbib]{revtex4}

\usepackage[colorlinks, linkcolor=blue, anchorcolor=blue, citecolor=blue]{hyperref}

\usepackage{graphicx}
\usepackage{color}
\usepackage{dcolumn}
\usepackage{bm}
\usepackage{slashed}
\usepackage{amsmath}
\usepackage{latexsym}
\usepackage{amssymb}
\usepackage{mathrsfs}
\usepackage{amsfonts}
\usepackage{url}
\allowdisplaybreaks

\begin{document}
\title{Extending Gibbons-Werner method to bound orbits of massive particles}

\author{Yang Huang}
\affiliation{Institute of Physics, Hunan University of Science and Technology, Xiangtan, 411201 Hunan, China}
\affiliation{Institute for Frontiers in Astronomy and Astrophysics, Beijing Normal University, Beijing 102206, China}
\affiliation{Department of Astronomy, Beijing Normal University, Beijing 100875, China}

\author{Bing Sun}
\affiliation{CAS Key Laboratory of Theoretical Physics, Institute of Theoretical Physics, Chinese Academy of Sciences, Beijing 100190, China}

\author{Zhoujian Cao}
\email[Corresponding author. \\ ]{zjcao@bnu.edu.cn}
\affiliation{Institute for Frontiers in Astronomy and Astrophysics, Beijing Normal University, Beijing 102206, China}
\affiliation{Department of Astronomy, Beijing Normal University, Beijing 100875, China}
\affiliation{School of Fundamental Physics and Mathematical Sciences, Hangzhou Institute for Advanced Study, UCAS, Hangzhou 310024, China}

\begin{abstract}
The Gibbons-Werner method for the gravitational deflection angle of unbound particles in static spherically symmetric spacetimes is based on Jacobi metric and Gauss-Bonnet theorem. When it is extended to bound massive particles, there exists two difficulties: (a) Bound orbits may overlap with themselves azimuthally. To extend the definition of deflection angle for unbound orbits to bound orbits, we divide the bound orbit into multiple segments such that each segment does not overlap with itself azimuthally and can be regarded as an unbound orbit. (b) The infinite region constructed for unbound orbits in Gibbons-Werner method is invalid for bound orbits, since the Jacobi metric of bound massive particles is singular at far region. To construct a suitable region for bound orbits, we adopt the generalized Gibbons-Werner method proposed in our last work [Huang and Cao, \href{https://journals.aps.org/prd/abstract/10.1103/PhysRevD.106.104043}{Phys. Rev. D 106, 104043 (2022)}], so that the unphysical region in Jacobi space is avoided. What's more, taking the Schwarzschild spacetime as an example, we show the details of the calculation and obtain an analytical expression of the deflection angle between two arbitrary points on the orbit.

\end{abstract}
\maketitle

\section{Introduction}\label{sec-introduction}
The deflection angle of particles moving in gravitational field is a significant quantity in relativistic astrophysics. The particle moving in a strong gravitational field can be treated as a test particle when its size is small enough comparing to the characteristic length of the system. Such orbits can be regarded as geodesics in background spacetimes. Whether the particle's orbit is bound or unbound is determined by the initial motion parameters such as energy and angular momentum.

The unbound orbit of massless particles (photons) has been deeply investigated. In 1919, Dyson, Eddington, and Davidson confirmed the deflection of light passing near the Sun \cite{dyson1920ix}, which laid the foundation of its usage in astrophysics and cosmology. Later the gravitational lensing based on the light deflection becomes one of the most important tools in the study of galaxies and clusters \cite{walsh19790957}, Hubble constant \cite{rhee1991estimate,koopmans2003hubble}, dark energy \cite{oguri2010gravitationally,treu2010strong}, dark matter \cite{aubourg1993evidence,alcock2000macho}, and extrasolar planets \cite{gaudi2008discovery,gould2010frequency}. The unbound orbit of massive particles has received less attention than that of massless particles since the rare astrophysical sources. Neutrons, Neutrinos, cosmic rays from stars and supernova, and theorized weakly interacting massive particles and Axions \cite{patla2013flux} may be the potential messenger, with which we can reveal the property of sources (e.g. supernova mechanism), lenses (e.g. mass, charge and angular momentum) and particles themselves \cite{liu2019constraining,accioly2004photon, bhadra2007testing,tsupko2014unbound,liu2016gravitational,he2016gravitational,he2017analytical,pang2019gravitational,li2019gravitational}.

The deflection angle is of great significance in the investigation of unbound orbits, thus people proposed various approaches to calculate it. For the photons, Gibbons and Werner introduced a geometric method to study the deflection angle of photons in static spherically symmetric (SSS) spacetimes in 2008 \cite{gibbons2008applications}, which is called Gibbons-Werner (GW) method later. By applying the Gauss-Bonnet theorem (GBT) to an infinite region in optical space, the GW method reveals the global properties of the deflection of light. In 2012, Werner extended this method to photons in stationary axially-symmetric (SAS) spacetimes with the help of Randers optical metric \cite{werner2012gravitational}. Henceforth, by applying the GW method, a series of works studying the deflection of photons with various condition in various background spacetimes emerged \cite{jusufi2016gravitational,jusufi2016light,ishihara2016gravitational,jusufi2017quantum,jusufi2017deflection,ishihara2017finite,jusufi2017light,jusufi2017rotating,jusufi2017effect,sakalli2017hawking,jusufi2017lightdeflection,goulart2017phantom,ono2017gravitomagnetic,jusufi2018effect,jusufi2018deflection,jusufi2018semiclassical,arakida2018light,ono2018deflection,jusufi2018gravitational97,jusufi2018conical,jusufi2018deflectionrotating,ovgun2018light,ovgun2018shadow,ovgun2018gravitational,jusufi2019gravitational,ovgun2019exact,ovgun2019weak,javed2019deflection,javed2019effect100,de2019weak,kumar2019shadow,jusufi2019light,javed2019effect,javed2020weak,javed2020weak135,javed2022weak,javed2022effect,mustafa2022shadows,gao2023microlensing,gao2023microlensingeffects,qiao2022weak}. For the massive particles, Crisnejo $et\ al.$ and Jusufi are respectively the ones firstly calculate the deflection angle in SSS \cite{crisnejo2018weak} and SAS \cite{jusufi2018gravitational} spacetimes with GW method. More works investigating the deflection of massive particles with GW method can be found in \cite{jusufi2019distinguishing,jusufi2019deflection,crisnejo2019gravitational,crisnejo2019higher,li2020gravitational,li2020finite,li2020thefinite,li2021kerrblack,li2021deflection,li2021kerr,huang2023finite}. 

In GW method, a suitable region which the GBT can be applied to must be constructed. All of the applications of the GW method to unbound orbits are based on an infinite region in previous literatures, except for \cite{takizawa2020gravitational} whose scheme is very complicated, \cite{li2020circular} which is only valid for a special case, and our recent work \cite{huang2022generalized} which introduces a generalized approach to construct the region freely and simplify the calculation.

Among the researchers studying the unbound orbits, some are no longer satisfied with the assumption that both the source and the observer are at infinity from the lensing object, since they want to explore the finite distance corrections to the deflection angle, and investigate the gravitational deflection in asymptotically nonflat spacetimes. To analyze the contribution of the cosmological constant to the gravitational lensing, in 2007, Rindler and Ishak proposed a definition of finite-distance deflection angle of light in a special situation where the lens, receiver, and source are aligned \cite{rindler2007contribution}. But this approach has been criticized by people \cite{park2008rigorous,sereno2009role,simpson2010lensing,arakida2012effect,ishihara2016gravitational}. In 2016, a finite-distance deflection angle was introduced by Ishihara $et\ al.$ with GW method \cite{ishihara2016gravitational}, and later was discussed by Crisnejo $et\ al.$\cite{crisnejo2019finite} and Takizawa $et\ al.$ \cite{takizawa2020gravitational}. In 2018, Arakida gave a definition of finite-distance deflection angle, however he compared geodesics which belong to two different spacetimes and did not present the mathematical justification \cite{arakida2018light}.

Up to now, only the finite-distance deflection angle defined by Ishihara $et\ al.$ \cite{ishihara2016gravitational} has been widely recognized. People usually adopt it in GW method to study the finite-distance deflection angle of massless and massive particles, the infinite-distance angle in GW method can also be deduced from it. As a broadly accepted definition, it can not only be applied to unbound orbits, but also to bound orbits.

Based on the model of bound orbits of massive particles, Einstein calculated the relativistic pericenter advance angle of Mercury \cite{einstein1915erklarung} which tested the general relativity successfully for the first time. The correction from general relativity for the bound orbit of massive particles is studied with great accuracy in \cite{ohta1973physically, ohta1974higher, ohta1974coordinate, damour2000poincare, blanchet2005structure}. The pericenter advance of the bound orbit is of great significance\cite{will2018new, tucker2019pericenter}, and the zoom-whirl orbit, as an extreme form of pericenter advance, also attracts people's attention \cite{glampedakis2002zoom, grossman2009dynamics, healy2009zoom, gold2013eccentric}. Observations on the astronomical phenomena that can be described by the model of bound massive particles have been performed for several decades. Since the first detection of stars around the supermassive black hole Sgr A* \cite{eckart1996observations,eckart1997stellar}, the number of detected S-stars increased with the improvement of instrumentation and analysis techniques \cite{peissker2020s62}. Recently, the star S2 has been followed through its pericenter passage using GRAVITY \cite{abuter2017first} at the Very Large Telescope Interferometer (VLTI) interferometer \cite{abuter2018detection, abuter2020detection}. In the foreseeable future, the Thirty Meter Telescope with higher angular resolution and deeper spectroscopy could provide better measurements to the stellar orbits \cite{sanders2013thirty,skidmore2015thirty}.

In this paper, (a) The definition of the finite-distance deflection angle given by Ishihara $et\ al.$ \cite{ishihara2016gravitational} is extended to bound orbits, so does its correspondence to observations. (b) By using the generalized GW method proposed by Huang and Cao \cite{huang2022generalized}, a finite region is chosen freely in the physically allowed area of the Jacobi space, and the formula for the deflection angle between two arbitrary points on trajectories is derived. Additionally, we show the calculation process of our scheme and formula in the Schwarzschild spacetime, and obtain an analytical result. 

The rest of this paper is organized as follows. In Sec.~\ref{sec-JacobiGaussAngle}, we review the Jacobi metric and GBT, and introduce the definition of finite-distance deflection angle proposed by Ishihara $et\ al.$ In Sec.~\ref{sec-bound}, we extend the finite-distance deflection angle for unbound orbits in GW method to bound orbits, and derive a formula of the deflection angle of bound massive particles in SSS spacetimes. In Sec.~\ref{sec-Schwarzschild}, the calculation of the deflection angle in Schwarzschild spacetime is performed. The spacetime signature ($-, +, +, +$) and the geometric units $G=c=1$ are used throughout the paper.

\section{Jacobi metric, GBT and deflection angle}
\label{sec-JacobiGaussAngle}

\subsection{Jacobi metric}
\label{subsec-JacobiMetric}
The geometry of the optical metric, also known as Fermat geometry or optical reference geometry \cite{werner2012gravitational}, was applied to the discussion of inertial forces in general relativity by Abramowicz $et\ al.$ \cite{abramowicz1988optical}, and thermal Green’s functions of black holes by Gibbons and Perry \cite{gibbons1993black}. With Fermat's principle \cite{perlick2006fermat,perlick2000ray}, the spatial projection of a lightlike geodesic corresponds to a geodesic of the optical geometry. Similarly, the Jacobi metric can be used to describe the timelike geodesics in static spacetimes \cite{gibbons2015jacobi}.

For a free massive particle moving in a static spacetime whose metric reads
\begin{equation}
    d s^{2}=g_{00}\left(\boldsymbol{x}\right) d t^{2}+g_{i j}\left(\boldsymbol{x}\right) d x^{i} d x^{j},
    \label{ds2}
\end{equation}
the corresponding Jacobi metric can be written as \cite{gibbons2015jacobi}
\begin{equation}
    d l^{2}=m^2 \left[\mathcal{E}^{2}+ g_{00}\left(\boldsymbol{x}\right)\right] \frac{g_{i j}\left(\boldsymbol{x}\right)}{-g_{00}\left(\boldsymbol{x}\right)} d x^{i} d x^{j},
    \label{dl2}
\end{equation}
where $\boldsymbol{x}$ is the spatial coordinate, the indices $i$ and $j$ both run from $1$ to $3$, $m$ and $\mathcal{E}$ denote the rest mass and the energy per unit rest mass for the particle, respectively. A timelike geodesic in the original four-dimensional static spacetime equiped with metric~\eqref{ds2} can be put in one-to-one correspondence with a geodesic in the three-dimensional Jacobi space equiped with metric~\eqref{dl2}.

Generally, if the spacetime is asymptotically flat, we have $-1 \leq g_{00}\left(\boldsymbol{x}\right) \leq 0$. If $\mathcal{E} \geq 1$, the Jacobi metric is positive definite. If $\mathcal{E} < 1$, which corresponds to bound orbits, the signature of Jacobi metric changes at large distances. What's more, there is a level set of $g_{00}\left(\boldsymbol{x}\right)$ on which $\mathcal{E}^2+g_{00}\left(\boldsymbol{x}\right)$ vanishes (i.e. the Jacobi metric becomes singular). From the perspective of Jacobi metric, this level set is a point-like conical singularity, and all geodesics have a turning point on or inside this level set \cite{gibbons2015jacobi}.

Furthermore, if the spacetime is SSS, Eq.~\eqref{ds2} will reduce to
\begin{equation}
    ds^2=g_{tt}\left(r\right) d t^{2}+g_{rr}\left(r\right) d r^{2}+ r^2 \left(d\theta^2+\sin^2\theta d\phi^2\right),
    \label{sss}
\end{equation}
and the corresponding Jacobi metric becomes (for notational simplicity we drop '$\left(r\right)$' for $g_{tt}\left(r\right)$ and $g_{rr}\left(r\right)$)
\begin{equation}
    d l^{2}=m^2 \left(\mathcal{E}^{2}+g_{tt}\right)\left[\frac{g_{rr}}{-g_{tt}} d r^{2}+\frac{r^2}{-g_{tt}}\left(d\theta^2+\sin^2\theta d\phi^2\right)\right].
    \label{JacobiMetric}
\end{equation}
Restricting our attention to the equatorial plane ($\theta=\pi/2$ and $d\theta=0$) without loss of generality, the Jacobi metric Eq.~\eqref{JacobiMetric} reduces to
\begin{equation}
    \begin{aligned}
    d l^{2}=& \alpha_{rr} dr^2 + \alpha_{\phi\phi} d\phi^2 \\
           =&m^2 \left(\mathcal{E}^{2} +g_{tt}\right)\left(\frac{g_{rr}}{-g_{tt}} d r^{2}+\frac{r^2}{-g_{tt}} d \phi^{2}\right),
    \end{aligned}
    \label{riemanDl2}
\end{equation}
with which we obtain a two-dimensional Riemann manifold corresponding to the equatorial plane of the Jacobi space of SSS spacetimes, and denote it as $M^{\left(\alpha 2\right)}$ for simplicity.

According to the cyclic coordinate $\phi$, we have a conserved quantity
\begin{equation}
    \mathcal{L}= m^2 \left(\mathcal{E}^{2} +g_{tt}\right)\frac{r^2}{-g_{tt}}\left(\frac{d \phi}{d l}\right),
    \label{dphidl}
\end{equation}
which represents the angular momentum per unit rest mass for the particle. Dividing both sides of Eq.~(\ref{riemanDl2}) by $d l^{2}$, then combining the result with Eq.~(\ref{dphidl}), the radial equation of the geodesics in $M^{\left(\alpha 2\right)}$ can be written as
\begin{equation}
    m^2\left(\mathcal{E}^{2} +g_{tt}\right)^{2} \frac{g_{rr}}{-g_{tt}}\left(\frac{d r}{d l}\right)^{2} = \mathcal{E}^{2}+g_{tt} \left( 1+\frac{\mathcal{L}^2}{r^2}\right).
    \label{drdl}
\end{equation}

\subsection{Gauss-Bonnet theorem}

As a fabulous result of differential geometry in the aspect of building the relation between the curvature properties of a Riemannian manifold and its topological structure, the GBT connects the Gaussian curvature integral of a compact and oriented even-dimensional manifold with its topological invariant, viz. the Euler characteristic.

\begin{figure}[!ht]
	\includegraphics[width=0.9\columnwidth]{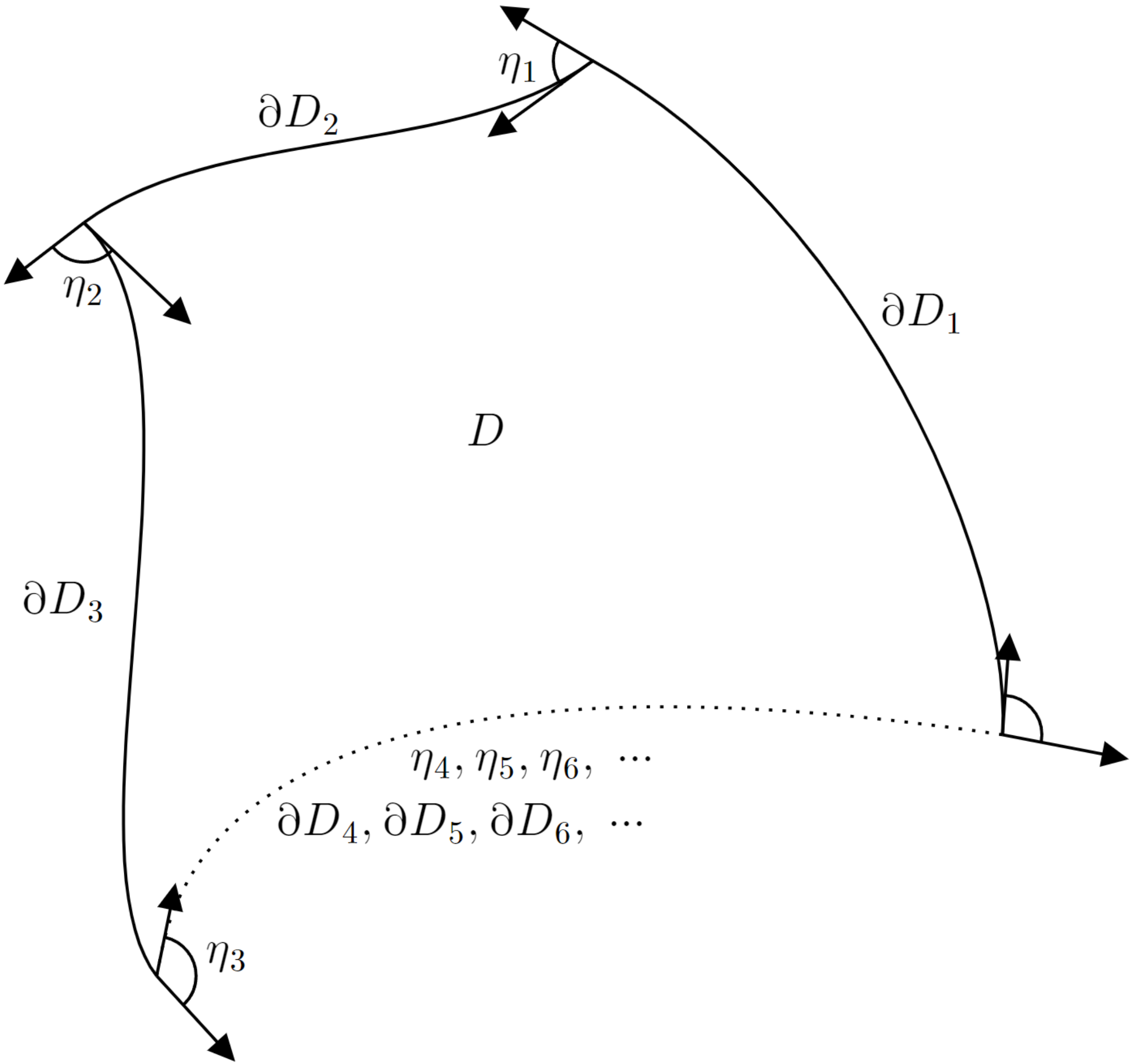}
	\caption{ Schematic figure for the GBT.}
	\label{fig-1}
\end{figure}

As shown in Fig.~\ref{fig-1}, suppose $D$ is a subset of a two-dimensional compact and oriented Riemannian manifold with Gaussian curvature $K$ and Euler characteristic number $\chi\left(D\right)$. Its boundary $\partial D = \bigcup\limits_i \partial D_i$ is a piecewise smooth curve. Then the GBT states
\begin{equation}
    \iint_{D} K d S+\sum_{i} \int_{\partial D_{i}} \kappa d l+\sum_{i} \eta_{i}=2 \pi \chi\left(D\right),
    \label{gbt}
\end{equation}
where $d S$ and $d l$ are respectively the area element and line element, $\eta_{i}$ denotes the exterior angle (or jump angle) at the $i$-th vertex in the sense of positive, $\kappa$ represents the geodesic curvature.

\subsection{Definition of finite-distance deflection angle for unbound particles}
\label{sec-definitionOfBendingAngle}
In this subsection, to review the definition of the finite-distance deflection angle for unbound particles, we (a) elaborate  Ishihara, Suzuki, Ono, Kitamura and Asada's original defining the deflection angle for photons whose source and observer are both assumed at finite distance \cite{ishihara2016gravitational}, (b) discuss the finite-distance deflection angle based on the observation of the lensed light, and (c) introduce people's successful applying the definition to not only massless particles but also massive particles.



\begin{figure}[!ht]
	\includegraphics[width=0.9\columnwidth]{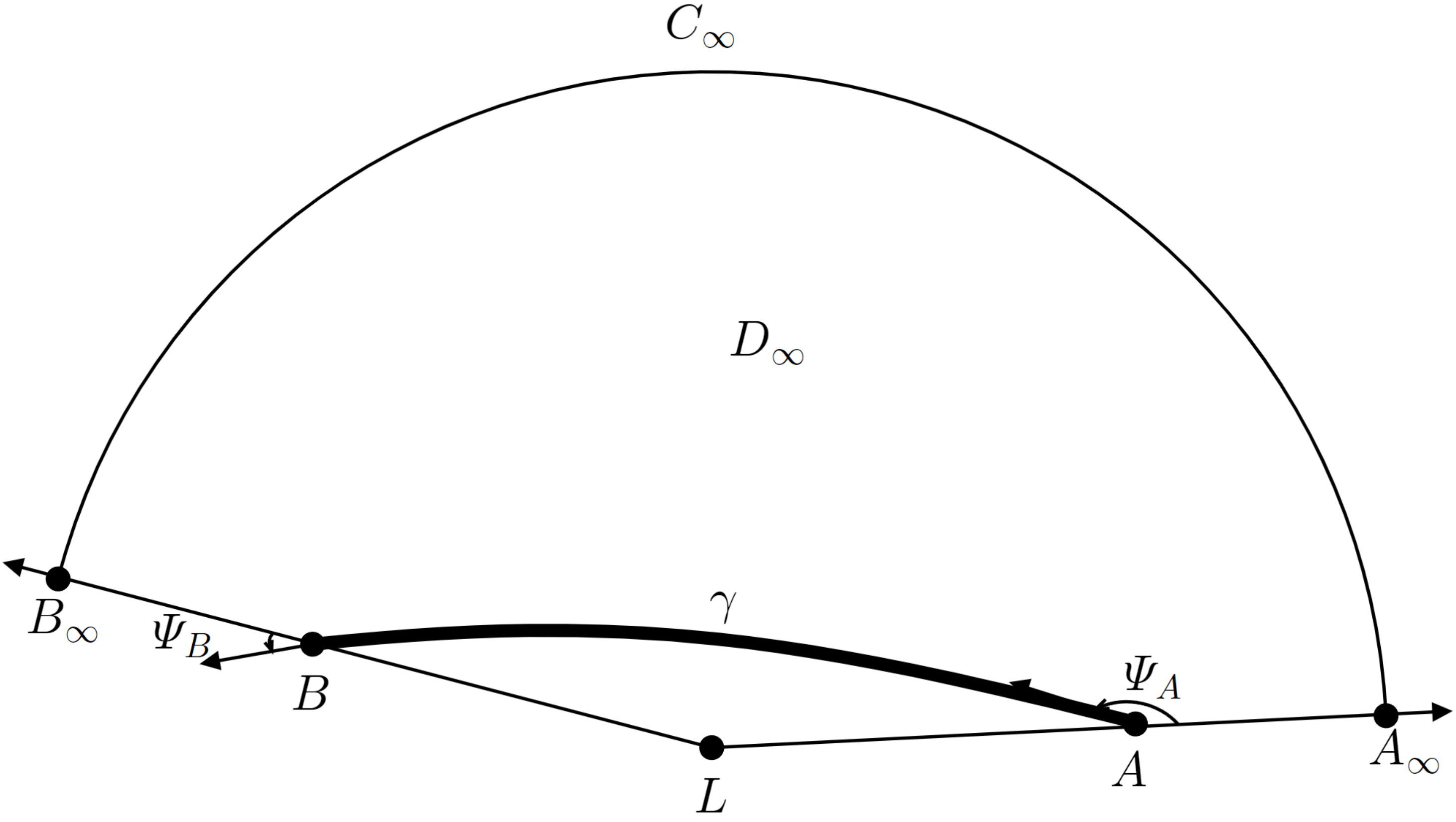}
	\caption{In the two-dimensional optical space corresponding to the metric~\eqref{opticalMetric}, $L$ is the lens, $\gamma$ is the trajectory of photons, $A$ and $B$ are two arbitrary points on $\gamma$, $\Psi_A$ and $\Psi_B$ are angles between the tangent vector along $\gamma$ and the outward radial direction measured at $A$ and $B$, respectively.}
	\label{fig-2}
\end{figure}

First, we give an introduction to the finite-distance deflection angle defined by Ishihara $et\ al.$ \cite{ishihara2016gravitational}. For the photons in the equatorial plane of an asymptotically flat SSS spacetime equiped with metric~\eqref{sss}, the optical metric states \cite{gibbons2008applications}
\begin{equation}
    dt^2 = \frac{g_{rr}}{-g_{tt}} dr^2 + \frac{r^2}{-g_{tt}} d\phi^2.
    \label{opticalMetric}
\end{equation}
As shown in Fig.~\ref{fig-2},
in the two-dimensional optical space corresponding to the metric~\eqref{opticalMetric}, $\gamma$ represents the photon's trajectory which is a geodesic, $L$ is the lens, $A$ and $B$ are two arbitrary points on $\gamma$, $\Psi_A$ and $\Psi_B$ are angles between the tangent vector along $\gamma$ (velocity direction) and the outward radial direction measured at $A$ and $B$, respectively. Then the finite-distance deflection angle along $\gamma$ between $A$ and $B$ is defined as \cite{ishihara2016gravitational}
\begin{equation}
    \delta_{BA} =  \Psi_B - \Psi_A + \phi_{BA},
    \label{deflectionangle}
\end{equation}
where $\phi_{BA}=\phi_B-\phi_A$ is the increment of the azimuthal position, $\phi_A$ denotes the longitude of $A$ measured from $L$, similarly the $\phi_B$. When both $A$ and $B$ approach infinity, namely $\Psi_A=\pi$ and $\Psi_B=0$, Eq.~\eqref{deflectionangle} reduces to the conventional infinite-distance deflection angle $\phi_{BA}-\pi$. 
To demonstrate that Eq.~\eqref{deflectionangle} is well defined, Ishihara $et\ al.$ apply the GBT to an infinite region $D_\infty = _B^{B_\infty}\square_A^{A_\infty}$, and obtain
\begin{equation}
    \begin{aligned}
    &\iint_{D_\infty} K dS + \int_{\overrightarrow{AA}_\infty} \kappa dl + \int_{C_\infty} \kappa dl + \int_{\overrightarrow{B_\infty B}} \kappa dl \\
    & + \int_{\overset{\curvearrowright}{BA}} \kappa dl + \eta_A + \eta_{A_\infty} + \eta_{B_\infty} + \eta_B = 2\pi\chi\left(D_\infty\right),
    \end{aligned}
    \label{GBT001}
\end{equation}
where $A_\infty$ is the intersection point of the outward radial curve $\overrightarrow{AA}_\infty$ and infinite circular arc $C_\infty=\overset{\curvearrowright}{A_\infty B_\infty}$, similarly the $B_\infty$. $\int_{C_\infty}\kappa dl = \int_{C_\infty} \left( 1/r_\infty\right)\cdot r_\infty d\phi=\phi_{BA}$ since $C_\infty$ is located at the asymptotic region, $\kappa\left( \overrightarrow{AA}_\infty\right)=\kappa\left( \overrightarrow{B_\infty B}\right)=0$, $\eta_A=\pi-\Psi_A$, $\eta_B=\Psi_B$, $\eta_{A_\infty}=\eta_{B_\infty}=\pi/2$. Then Eq.~\eqref{GBT001} becomes
\begin{equation}
    \delta_{BA} = -\iint_{D_\infty} K dS,
\end{equation}
which indicates the definition~\eqref{deflectionangle} is geometric invariant and well defined.

\begin{figure}[!ht]
	\includegraphics[width=0.9\columnwidth]{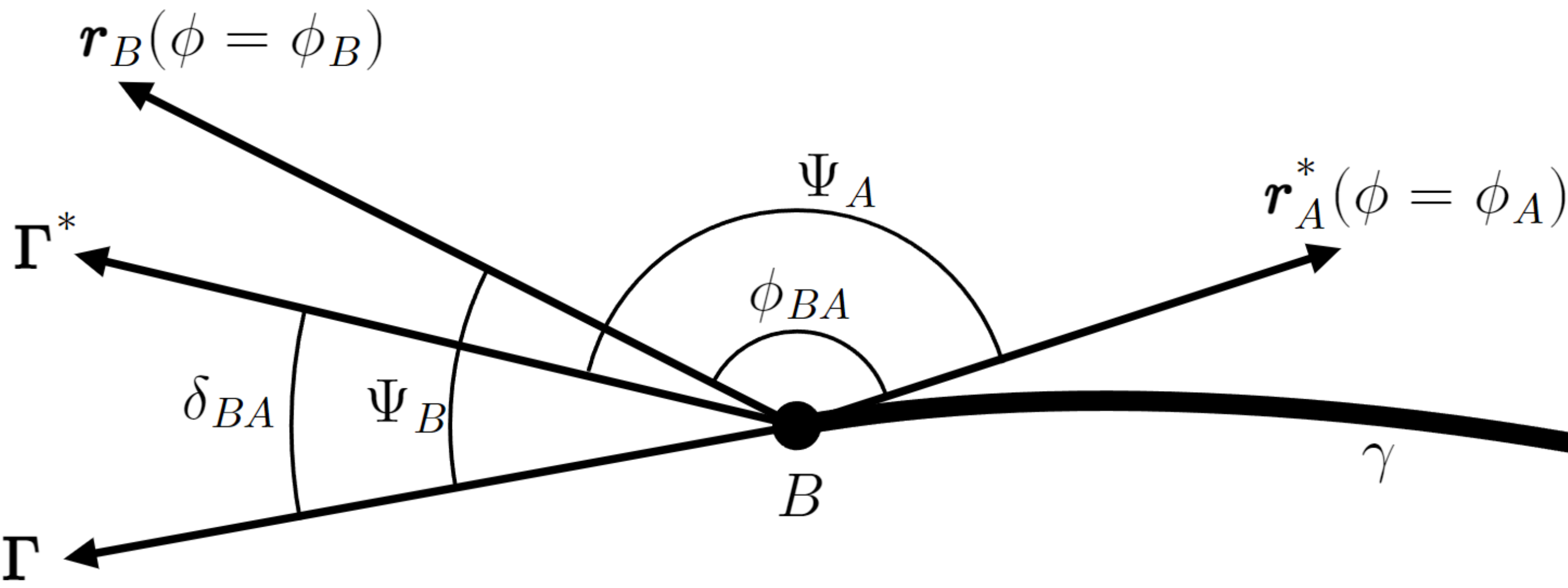}
	\caption{Focusing on the observer $B$ in Fig.~\ref{fig-2}, $\boldsymbol{\Gamma}$ is the real direction of light at $B$, $\boldsymbol{\Gamma}^*$ is the fiducial source direction, $\boldsymbol{r}_B$ is the radial direction at $B$, $\boldsymbol{r}_A^*$ is the fiducial radial direction.}
	\label{fig-temp1}
\end{figure}

Second, we present the understanding of the definition~\eqref{deflectionangle} from the perspective of the observation. In 2020, Takizawa, Ono and Asada interpreted Eq.~\eqref{deflectionangle} as the angle between the real light direction (the direction of light rays coming from the lensed image) and the fiducial source direction (the direction of the light rays coming from the unlensed source) at the position of the observer \cite{takizawa2020gravitational}. As show in Fig.~\ref{fig-temp1},
focusing on the observer $B$ in Fig.~\ref{fig-2}, $\boldsymbol{\Gamma}$ is the direction of the tangent vector along $\gamma$ at $B$, i.e. velocity direction at $B$, it corresponds to the real direction of the light. $\boldsymbol{\Gamma}^*$ is the fiducial source direction of the light. From the sight of the observer, the deflection angle $\delta_{BA}$ should be the deviation between the real direction and the fiducial direction. This scheme imitates that of Eddington's observation to the light bended by the Sun, in which a comparison is made between the observed (lensed) image direction and the intrinsic fiducial (unlensed) source direction to obtain the deflection angle \cite{dyson1920ix}. Now the problem becomes the determination of the fiducial source direction $\boldsymbol{\Gamma}^*$. One can define the fiducial radial direction $\boldsymbol{r}^*_A$ such that the angle between $\boldsymbol{r}_B$ (the radial direction at the observer) and $\boldsymbol{r}^*_A$ is equal to $\phi_{BA}$ which can be obtain from the ephemeris or other channels. Thus $\boldsymbol{\Gamma}^*$ is determined by rotating the fiducial radial direction $\boldsymbol{r}^*_A$ with angle $\Psi_A$ which can be measured by observers \cite{ono2019effects}. According to Fig.~\ref{fig-temp1}, we have
\begin{equation}
    \Psi_B - \delta_{BA} = \Psi_A - \phi_{BA}
\end{equation}
at the position of the observer, i.e. $\delta_{BA} = \Psi_B-\Psi_A+\phi_{BA}$ from the perspective of observers. What's more, the gravitational deflection of orbits can be regarded as the result of the combination of two factors---the velocity direction $\Psi$ (with respect to the radial direction) and the azimuthal position $\phi$. Consequently we can recast Eq.~\eqref{deflectionangle} into the form
\begin{equation}
    \delta_{BA} = \delta_{BA}^{\Psi} + \delta_{BA}^\phi,
    \label{deltapsiphi}
\end{equation}
where $\delta_{BA}^{\Psi}=\Psi_B-\Psi_A$ and $\delta_{BA}^\phi=\phi_B-\phi_A$. More detailed analysis can be found in \cite{takizawa2020gravitational}. 

Third, we give a brief summary of the application of the definition~\eqref{deflectionangle}, which has been extended to unbound orbits of massless and massive particles in various asymptotically flat and nonflat spacetimes. Specifically, Ono $et\ al.$ studied the deflection angle of photons with finite-distance corrections in SAS and asymptotically flat spacetimes especially the Kerr black hole \cite{ono2017gravitomagnetic} and rotating Teo wormhole \cite{ono2018deflection}, the case of asymptotically nonflat spacetimes is also investigated in \cite{ono2019deflection}. Haroon $et\ al.$ analyzed the deflection angle of photons for rotating black holes in perfect fluid dark matter with a cosmological constant \cite{haroon2019shadow}. Kumar $et\ al.$ discussed the weak gravitational lensing of the charged rotating regular black hole (a generalized Kerr-Newman black hole with a regular origin) \cite{kumar2019shadow}. Li and his collaborators studied the finite-distance gravitational deflection of neutral massive particles in SAS and asymptotically flat spacetimes \cite{li2020thefinite} and Kerr-like black hole in the bumblebee gravity model \cite{li2020finite}, the deflection of charged massive particles by a four-dimensional charged Einstein-Gauss-Bonnet black hole is also investigated by them \cite{li2021deflection}. More related works can been found in \cite{crisnejo2019finite,ono2019effects,li2021kerrblack,takizawa2020gravitational,li2020circular,huang2023finite}.

\section{Deflection angle of bound massive particles in GW method}
\label{sec-bound}

\subsection{Extending the definition of deflection angle for unbound orbits to bound orbits}
The definition~\eqref{deflectionangle} is proposed for and is usually applied to the unbound orbits which do not overlap with themselves azimuthally (with respect to the azimuthal coordinate $\phi$). To calculate the deflection angle along the bound orbit, we carefully extend Eq.~\eqref{deflectionangle} to the bound case by dividing the orbit into multiple segments such that each segment does not overlap with itself azimuthally.

\begin{figure}[!ht]
	\includegraphics[width=0.9\columnwidth]{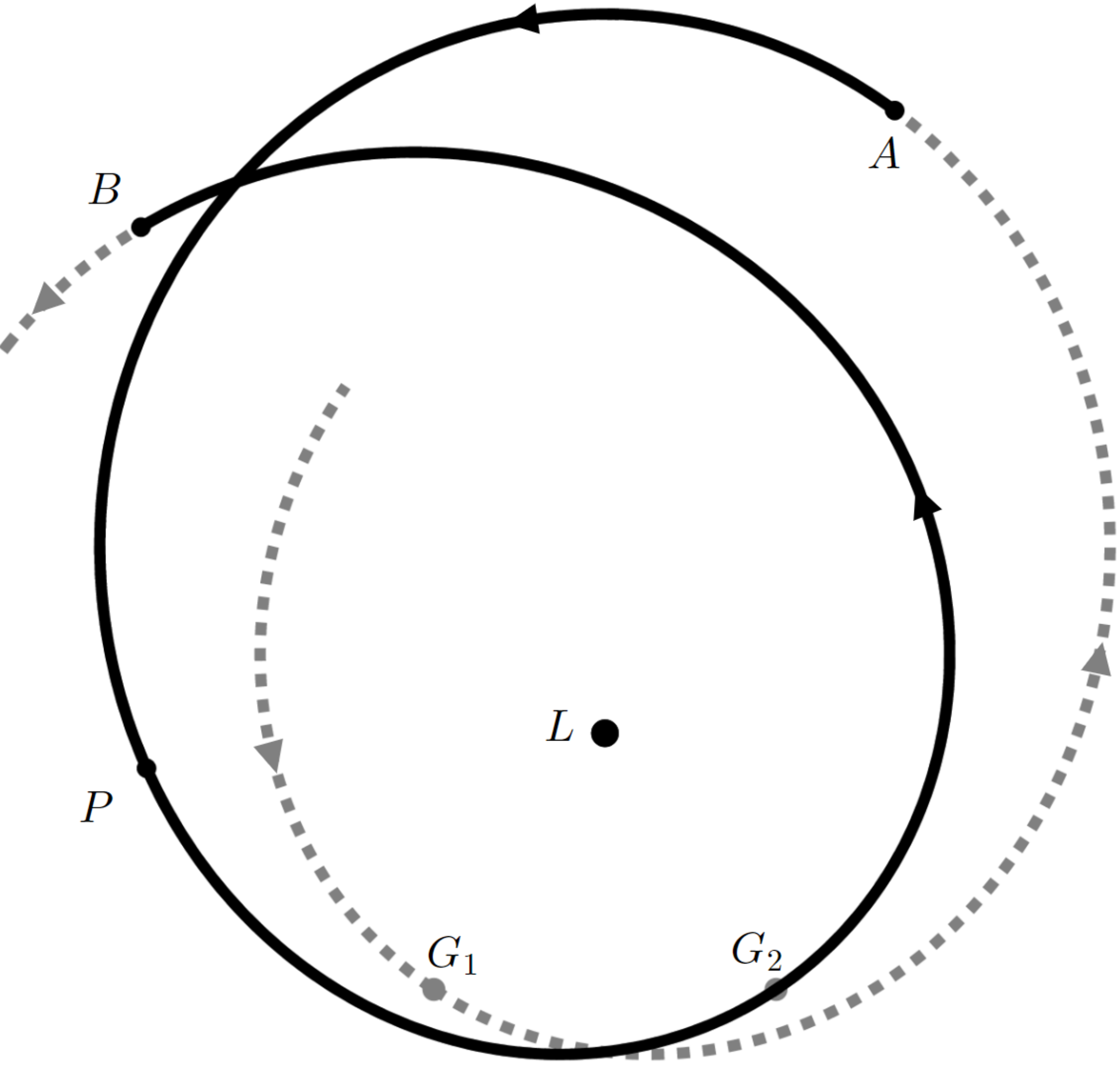}
	\caption{The bound trajectory of massive particles in $M^{\left(\alpha 2\right)}$.}
	\label{fig-4}
\end{figure}

Consider the bound trajectory of massive particles moving in $M^{\left(\alpha 2\right)}$, which is the two-dimensional Riemann manifold corresponding to the equatorial plane of the Jacobi space of SSS spacetimes as mentioned in Sec.~\ref{subsec-JacobiMetric}. As shown in Fig.~\ref{fig-4},
$L$ represents the black hole, $G_1$ and $G_2$ are two adjacent pericenters, $A$ and $B$ are two arbitrary points on the bound trajectory. $P$ is an arbitrary auxiliary point on the trajectory between $A$ and $B$ such that $\overset{\curvearrowright}{AP}$ and $\overset{\curvearrowright}{PB}$ do not overlap with themselves azimuthally. According to the definition~\eqref{deflectionangle}, we have
\begin{equation}
    \begin{aligned}
    &\delta_{PA}=\Psi_P- \Psi_A + \phi_{PA}, \\
    &\delta_{BP}=\Psi_B- \Psi_P + \phi_{BP}.
\end{aligned}
\end{equation}
Then the deflection angle along the trajectory from $A$ to $B$ can be written as
\begin{align}
    \delta_{BA} = & \delta_{PA}+\delta_{BP}  \label{boundDEF} \\
                = & \Psi_B-\Psi_A+\phi_{BA}. \label{boundangle2}
\end{align}
For the trajectory with more windings, the above formula still holds which can be demonstrated by dividing the trajectory into more segments.

It should be noted that, we select the counterclockwise direction as the rotation direction of the particle, which does not affect our discussion. Considering the cyclicity of the azimuthal coordinate of bound trajectories, in this paper we assume that the azimuthal coordinate monotonically increases with the counterclockwise rotation of the particle. Consequently, $\phi_{BA}$ and $\delta_{BA}$ may be greater than $2\pi$.

\subsection{Correspondence to observations}
\label{sec-observation}
Similar to the discussion of definition~\eqref{deflectionangle} for unbound photons in the second paragraph of Sec.~\ref{sec-definitionOfBendingAngle} (also Sec.~II.B of \cite{takizawa2020gravitational}), we give an understanding of that for bound massive particles by taking the Mercury as an example.

The ephemeris of the Mercury can be obtained by taking into account not only the gravitational interaction between the Mercury and the Sun but also effects from all other planets, the Earth’s moon, and 300 of the most massive asteroids, as well as interactions between Earth and Moon caused by nonsphericity and tidal effects. In fact, to provide the observational pericenter advance angle of the inner planets Mercury, Venus, Earth, and Mars for Taylor and Wheeler (\S 10.8, \cite{taylor2000exploring}), Myles Standish, the Principal Member of the Technical Staff at Jet Propulsion Laboratory (JPL), calculated orbits of the four inner planets over four centuries (from A.D. 1800 to A.D. 2200) by using the numerical integration program of the Solar System Data Processing System. For each inner planet, Standish worked out two types of orbit whose difference is whether the relativistic effect is considered, and obtained the pericenter advance angle accounted for only by general relativity through comparing the result from these two types of orbit.

JPL is devoted to obtaining the positions and velocities of the major bodies in the solar system as precisely as possible. For the major planets and the moon, a huge database is maintained by JPL and the Solar System Data Processing System mentioned before refers to a set of powerful computer programs \cite{jpl}. Additionally, similar tasks are also performed by the Paris Observatory \cite{inpop} and the Institute of Applied Astronomy of the Russian Academy of Sciences \cite{epm}.

Assuming the trajectory of the Mercury can be illustrated by Fig.~\ref{fig-4}, $\overset{\curvearrowright}{AB}$ is the trajectory segment we concerned. Then we can apply Fig.~\ref{fig-temp1} to point $B$ in Fig.~\ref{fig-4}, the relevant quantities at $B$ such as $\Psi_B$ and $\phi_B$ can be obtained from the ephemeris, although the $B$ in Fig.~\ref{fig-4} is not the observer's position like that in Fig.~\ref{fig-temp1}. More over, $\Psi_A$ and $\phi_A$ can also be obtained from the ephemeris. Thus one can derive the observed $\delta_{BA}$. It should be remarked that in the actual calculation, the $\delta_{BA}$ must be computed based on considerable number of cycles or considerable period of time e.g. several Earth-centuries so that the statistical error can be reduced. Subtracting the result of nonrelativistic orbits from that of relativistic orbits, the $\delta_{BA}$ accounted for only by general relativity will be derived, and can be described with respect to a period of time such as '$\delta_{BA}$ per Earth-century'.

Since the radial coordinate of the trajectory in Fig.~\ref{fig-4} is periodic, it can be expressed as
\begin{equation}
    r = \frac{p}{1+e\cos\xi}, \quad \xi\in \left(-\infty, \infty\right),
    \label{rxi}
\end{equation}
where $e$ and $p$ are respectively the eccentricity and semi-latus rectum of the trajectory. Without loss of generality, we assume that the particle starts at the pericenter $G_1$ when $\xi=0$, then moves in to the apocenter when $\xi=\pi$, and arrive the next pericenter $G_2$ when $\xi=2\pi$. What's more, we use $\xi_A$ and $\xi_B$ to identify points $A$ and $B$, respectively. Then according to definition~\eqref{deflectionangle}, we have
\begin{equation}
    \begin{aligned}
    \delta_{BA} = &\Psi\left(\xi= 2n\pi+\xi_B+2\pi\right)-\Psi\left(\xi=2n\pi+\xi_A\right)+\\
    &\phi\left(\xi=2n\pi+\xi_B+2\pi \right)-\phi\left(\xi= 2n\pi+\xi_A \right),
    \end{aligned}
\end{equation}
where $n=0,\pm 1,\pm 2,\cdots$. In the actual calculation of the deflection angle from observations, for the orbit with vast number of cycles or a long period of time, one can use enough number of $n$ to cover all of the data from the ephemeris for the sake of reducing the statistical error.

It seems that the $\delta_{BA}$ is similar to the pericenter advance angle $\Delta\omega$. In general, $\Delta\omega$ only involves the azimuthal position of two special points---a pericenter and its adjacent pericenter, namely $\Delta\omega=\phi\left(\xi=2n\pi+2\pi \right)- \phi\left(\xi=2n\pi\right)-2\pi$. While $\delta_{BA}$ involves not only the azimuthal position but also the velocity direction, in addition points $A$ and $B$ can be chosen freely. It can be roughly concluded that $\delta_{BA}$ can encode more information than $\Delta\omega$. In fact, the pericenter advance angle (plus $2\pi$) can be regarded as the deflection angle between two adjacent pericenters, since for any pericenter the $\Psi$ is $\pi/2$ and does not contribute to the deflection angle.

\subsection{Formula of deflection angle for bound orbits with generalized GW method}

According to Eq.~\eqref{JacobiMetric}, in the Jacobi space of SSS spacetimes, only the region satisfying $\mathcal{E}^2+g_{tt}>0$ is nonsingular and can be treated as a Riemannian space. We denote the critical radial coordinate of the nonsingular region by $r_{cri}$. For example, consider the Schwarzschild spacetime for which $g_{tt} =-\left(1-2M/r\right)$, the energy of unit rest mass for bound massive particles must satisfy $\mathcal{E}<1$, thus $r_{cri}=2M/\left(1-\mathcal{E}^2\right)$. The region $r>r_{cri}$ is singular, thus the region that the GBT can be applied to is confined by $r<r_{cri}$. If the event horizon $r=2M$ is further considered, the bound massive particles in Jacobi space of Schwarzschild spacetime will be confined in the region $r\in\left(2M, 2M/\left(1-\mathcal{E}^2\right)\right)$.

\begin{figure}[!ht]
	\includegraphics[width=0.9\columnwidth]{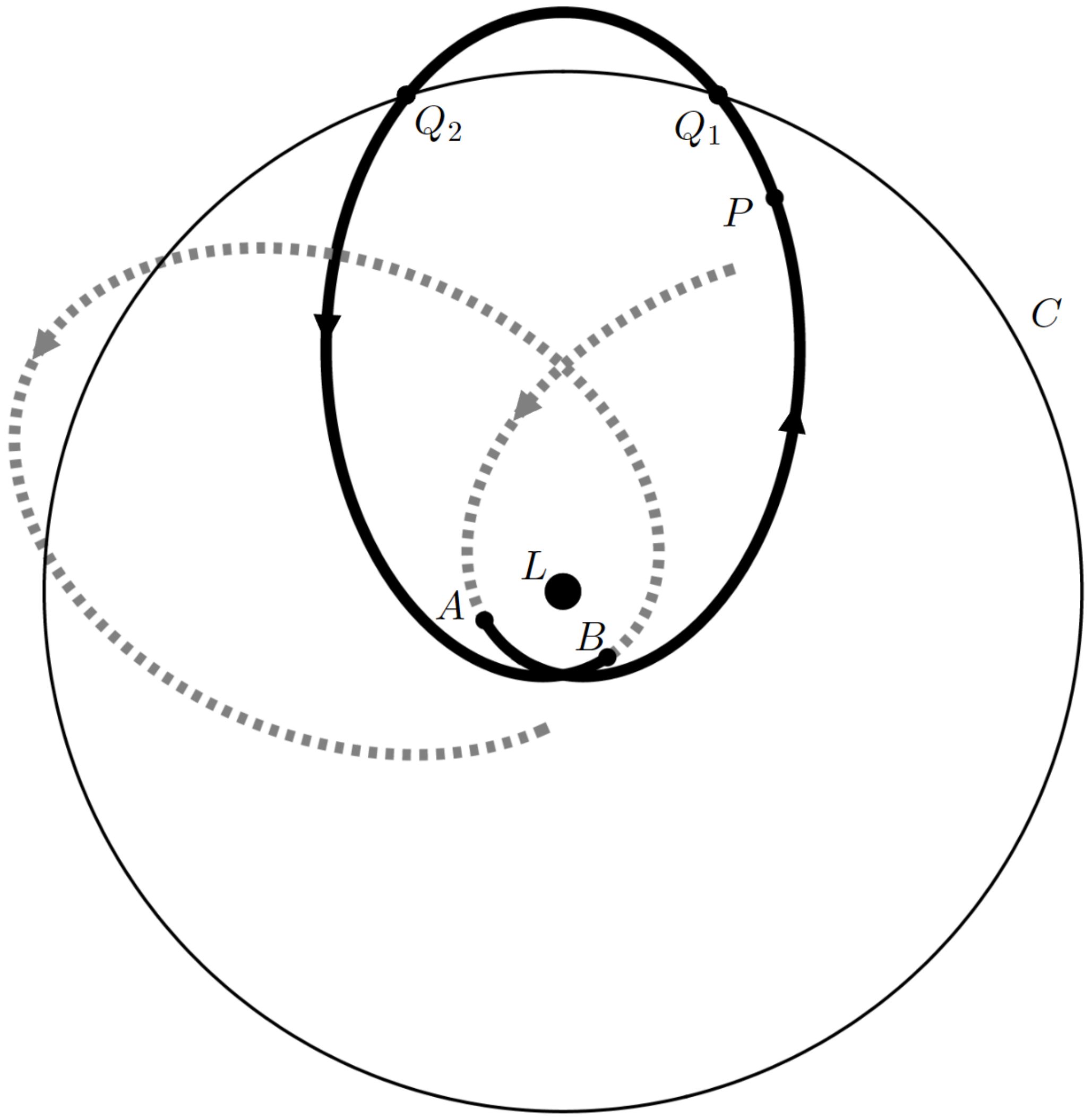}
	\caption{In $M^{\left(\alpha 2\right)}$, $\overset{\curvearrowright}{AB}$ is a segment of the trajectory of bound massive particles, $C$ denotes an auxiliary circle.}
	\label{fig-5}
\end{figure}

As shown in Fig.~\ref{fig-5},
in $M^{\left(\alpha 2\right)}$, $L$ denotes the black hole, $\overset{\curvearrowright}{AB}$ is a segment of the trajectory of bound massive particles. We divide $\overset{\curvearrowright}{AB}$ into two segments $\overset{\curvearrowright}{AP}$ and $\overset{\curvearrowright}{PB}$ which do not overlap with themselves azimuthally. $C$ is an auxiliary circle centered at $L$ with $r=r_0$, and being in the physical allowed region is the only requirement for $C$ or $r_0$. The auxiliary circle we choose in Fig.~\ref{fig-5} intersects with the trajectory, specifically $C$ intersects with $\overset{\curvearrowright}{AB}$ at $Q_1$ and $Q_2$.

\begin{figure}[!ht]
	\includegraphics[width=0.9\columnwidth]{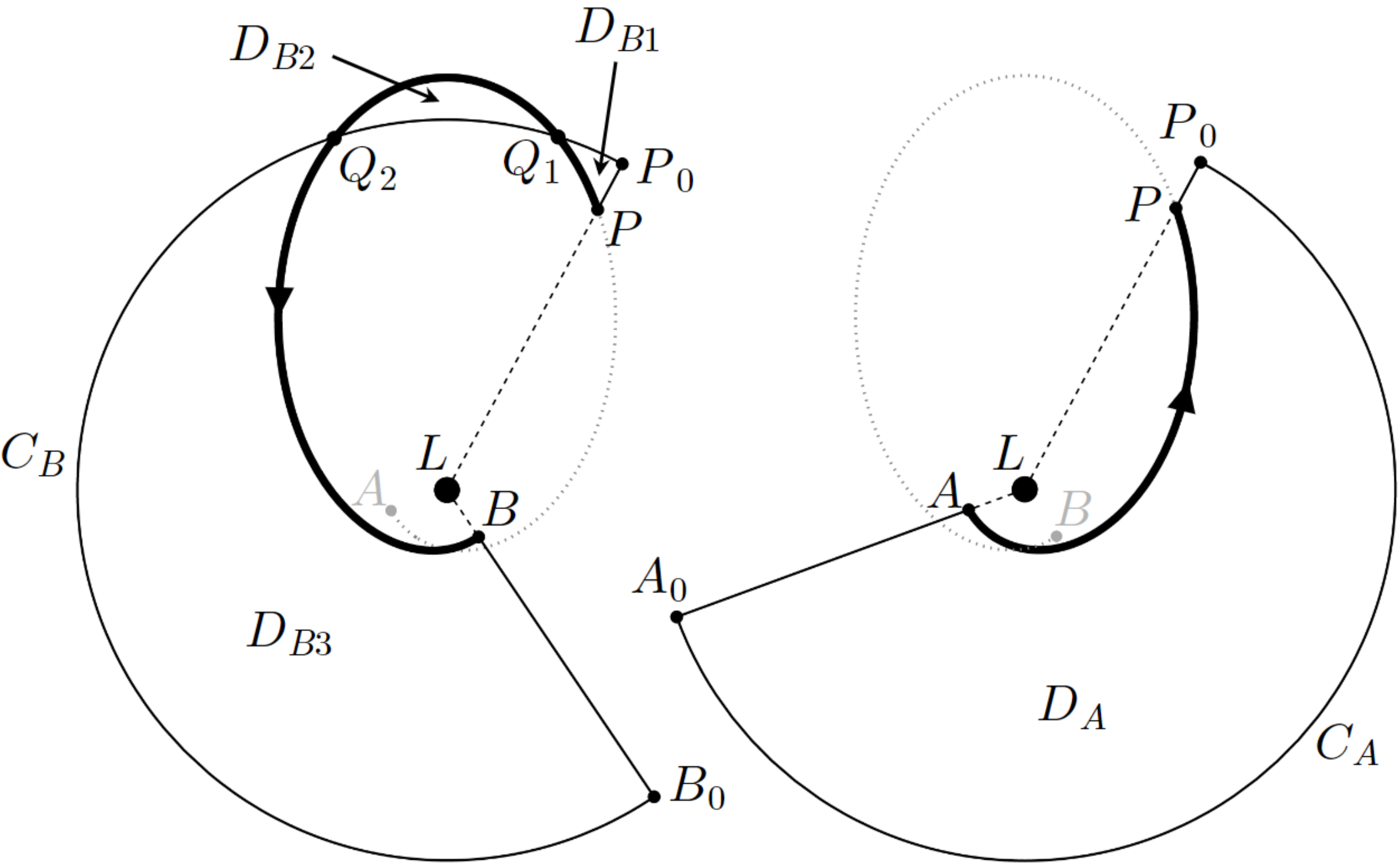}
	\caption{The trajectory segments $\overset{\curvearrowright}{AP}$ and $\overset{\curvearrowright}{PB}$ in Fig.~\ref{fig-5} are separated to have a better presentation. The auxiliary circular arc $C_{A}=\overset{\curvearrowright}{A_0 P_0}$, centered at $L$ with $r=r_0$, intersects with two outward radial geodesics passing through $A$ and $P$ at $A_0$ and $P_0$, respectively. The auxiliary circular arc $C_{B}=\overset{\curvearrowright}{P_0 B_0}$, centered at $L$ with $r=r_0$, intersects with two outward radial geodesics passing through $P$ and $B$ at $P_0$ and $B_0$, respectively. }
	\label{fig-6}
\end{figure}

As shown in Fig.~\ref{fig-6},
$\overset{\curvearrowright}{AP}$ and $\overset{\curvearrowright}{PB}$ are separated to have a better presentation. The auxiliary circular arc $C_{A}=\overset{\curvearrowright}{A_0 P_0}$, centered at $L$ with $r=r_0$, intersects with two outward radial geodesics passing through $A$ and $P$ at $A_0$ and $P_0$, respectively. Thus we get a quadrilateral region $D_{A} = _{A_0}^A\square_{P_0}^P$. According to the generalized GW method in \cite{huang2022generalized}, applying the GBT to $D_A$ yields
\begin{equation}
    \delta_{PA} = \int_{\phi_A}^{\phi_P} \left[1+H\left(r_\gamma \right) \right] d \phi,
    \label{deltaPA}
\end{equation}
where $r_\gamma$ is the radial coordinate of the trajectory, and
\begin{equation}
    H\left( r\right) = -\frac{\alpha_{\phi\phi,r}}{2\sqrt{\alpha}}.
    \label{Hr}
\end{equation}
$\alpha$ denotes the determinant of the metric of $M^{\left(\alpha 2\right)}$ (Eq.~\eqref{riemanDl2}). The auxiliary circular arc $C_{B}=\overset{\curvearrowright}{P_0 B_0}$, centered at $L$ with $r=r_0$, intersects with two outward radial geodesics passing through $P$ and $B$ at $P_0$ and $B_0$, respectively. Thus we get three regions: the triangle $D_{B1} = _{Q_1}\triangle^{P_0}_P$, the digon $D_{B2}$ with vertexes $Q_1$ and $Q_2$, and the triangle $D_{B3}=_{B_0}\triangle^{Q_2}_{B}$. According to the generalized GW method in \cite{huang2022generalized}, applying the GBT to $D_{B1}$, $D_{B2}$, and $D_{B3}$ yields
\begin{equation}
    \delta_{BP} = \int_{\phi_P}^{\phi_B} \left[1+H\left(r_\gamma \right) \right] d \phi.
    \label{deltaBP}
\end{equation}
Substituting Eqs.~\eqref{deltaPA} and \eqref{deltaBP} into Eq.~\eqref{boundDEF} leads to
\begin{equation}
    \delta_{BA} = \int_{\phi_A}^{\phi_B} \left[ 1+ H\left(r_\gamma\right) \right] d\phi,
    \label{deltaAB}
\end{equation}
which holds regardless of whether $r_0<r_\gamma^{min}$, $r_\gamma^{min}\le r_0 \le r_\gamma^{max}$, or $r_0>r_\gamma^{max}$ according to the discussion in \cite{huang2022generalized}, $r_\gamma^{min}$ and $r_\gamma^{max}$ are the minimum and the maximum of the radial coordinate of the trajectory, respectively. Eq.~\eqref{deltaAB} can also be derived based on Eq.~\eqref{boundangle2} by applying GBT to relevant regions ($D_A$, $D_{B1}$, $D_{B2}$, and $D_{B3}$) and simplifying formulas with the scheme of generalized GW method. If the trajectory segment we concerned intersects with the auxiliary circle at more points, Eq.~\eqref{deltaAB} still holds although it should be derived by constructing more regions (digon, triangle, and quadrilateral).

The flat spacetime can be looked as the zeroth-order counterpart of a curved spacetime with the post-Newtonian approximation, so does the relevant quantity. According to Eq.~\eqref{Hr}, the $H(r)$ in flat spacetime is $-1$ ($\alpha_{rr}=1$ and $\alpha_{\phi\phi}=r^2$), i.e. the zeroth-order term of $1+H\left( r\right)$ vanishes. As a consequence, by using Eq.~\eqref{deltaAB} one can extract the $\left(N+1 \right)$th-order deflection angle from the $N$th-order orbit.


\section{Deflection angle of bound massive particles in Schwarzschild spacetime}
\label{sec-Schwarzschild}
In this section, taking the Schwarzschild spacetime as an example, the calculation process of the formula~\eqref{deltaAB} is presented, and the deflection angle between two arbitrary points on the bound trajectory of massive particles is obtained and analyzed.

\subsection{Motion of bound massive particles}
The metric of Schwarzschild spacetime states
\begin{equation}
    ds^2= -\left( 1-\frac{2M}{r}\right) dt^2+\frac{1}{1-\frac{2M}{r}} dr^2 + r^2 d\theta^2 + r^2 \sin^2\theta d\phi^2.
\end{equation}
Without loss of generality, we focus on the massive particles moving in the equatorial plane. Substituting the above metric into Eq.~\eqref{riemanDl2} brings about the metric of the corresponding $M^{\left(\alpha 2\right)}$
\begin{equation}
    d l^{2}= m^2\left[\mathcal{E}^{2}-\left(1-\frac{2 M}{r}\right)\right] \left[\frac{d r^{2}}{\left(1-\frac{2 M}{r}\right)^{2}}+\frac{r^{2} d \phi^{2}}{1-\frac{2 M}{r}}\right].
    \label{JacobiMetricSch}
\end{equation}

According to Eqs.~\eqref{dphidl} and \eqref{drdl}, we obtain the equations of geodesic motion 
\begin{equation}
    \mathcal{L}= m^2 \left( \frac{\mathcal{E}^{2}}{ 1-\frac{2M}{r} } - 1 \right) r^2 \left(\frac{d \phi}{d l}\right),
    \label{dphids1}
\end{equation}
\begin{equation}
    m^2\left( \frac{\mathcal{E}^2}{ {1-\frac{2M}{r}} } - 1 \right)^2 \left( \frac{dr}{dl} \right)^2 = \mathcal{E}^{2}-\left(1-\frac{2M}{r} \right) \left( 1+\frac{\mathcal{L}^2}{r^2}\right).
    \label{drdls2}
\end{equation}
Let the right side of Eq.~(\ref{drdls2}) equal to zero, we obtain
\begin{equation}
    0 = \mathcal{E}^{2}-\left(1-\frac{2M}{r}\right) \left( 1+\frac{\mathcal{L}^2}{r^2}\right) ,
    \label{hr2}
\end{equation}
which can be recast into the form
\begin{equation}
    0 = 2M\mathcal{L}^2\left(\frac{1}{r}-\frac{1}{r_1}\right)\left(\frac{1}{r}-\frac{1}{r_2}\right)\left(\frac{1}{r}-\frac{1}{r_3}\right).
\label{hr2a}
\end{equation}
Obviously, $r_1$, $r_2$ and $r_3$ are roots of Eq.~\eqref{hr2}, also roots of equation $dr/d l = 0$. For the bound orbits of massive particles, we set (Sec.~5.6.3 of \cite{poisson2014gravity})
\begin{equation}
    r_1 = \frac{p}{1+e}, r_2=\frac{p}{1-e}, r_3=\frac{2Mp}{p-4M},\left(0<e<1\right),
\label{r1r2r3}
\end{equation}
where $e$ and $p$ are respectively the eccentricity and semi-latus rectum of the orbit, and are constant like the energy and the angular momentum. $r_1$ and $r_2$ correspond to the inner and outer turning points (pericenter and apocenter) of the orbit, respectively. 
Substituting Eq.~\eqref{r1r2r3} into Eq.~\eqref{hr2a}, and comparing the result with Eq.~\eqref{hr2}, $\mathcal{E}$ and $\mathcal{L}$ can be expressed in terms of $e$ and $p$
\begin{equation}
    \begin{aligned}
		\mathcal{E}^2 & = 1 -\frac{  M\left(1-e^{2}\right) }{ p} \frac{ p-4 M }{ p-\left(3+e^2\right) M }, \\
		\mathcal{L}^{2} & =   \frac{M p^2}{p-\left(3+e^{2}\right) M}. \\
	\end{aligned}
	\label{EL}
\end{equation}

We adopt the change of variable $r=p/\left(1+e\cos\xi \right)$ (Eq.~\eqref{rxi}) to describe the periodic motion of bound massive particles moving in the equatorial plane of Schwarzschild spacetime. According to Eqs.~\eqref{rxi}, \eqref{dphids1}, \eqref{drdls2}, and \eqref{EL}, we obtain
\begin{align}
	\frac{d \phi}{d \xi} = \pm \frac{1}{ \sqrt { 1-\frac{2 M}{p}\left(3+e \cos \xi\right) } }.
	\label{dphidchi}
\end{align}
The '$\pm$' in above expression represents two different rotation directions of massive particles, which is unimportant for our calculation. Without loss of generality, the plus sign of Eq.~\eqref{dphidchi} is selected, which means the particles rotate counterclockwise and $\phi$ monotonically increases as $\xi$ increases.

\subsection{The deflection angle}
In this subsection, we calculate the deflection angle between two arbitrary points on the trajectory with Eq.~\eqref{deltaAB}. Substituting Eq.~\eqref{JacobiMetricSch}, the metric of the equatorial plane of Jacobi space corresponding to massive particles in Schwarzschild spacetime, into Eq.~\eqref{Hr}, we derive
\begin{equation}
    H\left(r\right)=\frac{r \mathcal{E}^2 \left(3 M-r\right)+\left(r-2 M\right)^2}{\left[2 M+r \left(\mathcal{E}^2-1\right)\right]\sqrt{r \left(r-2 M\right)} }.
\end{equation}
Then using Eq.~\eqref{rxi} ($r_\gamma=p/\left(1+e \cos\xi\right)$) and Eq.~\eqref{dphidchi}, we get the indefinite integral
\begin{equation}
    \begin{aligned}
    \int \left[ 1+ H\left(r_\gamma\right) \right] d\phi = & \int \left[ 1+ H\left(r_\gamma\right) \right] \frac{d\phi}{d\xi}d\xi \\ =& \Omega\left(\xi\right)+Const,
    \label{schdelta}
\end{aligned}
\end{equation}
in which

\begin{widetext}
    \begin{equation}
        \begin{aligned}
        \Omega\left(\xi\right) =&  \frac{ 2 e p }{\left(e^{2}-1\right) \sqrt{\mathcal{B}\mathcal{D}}} i
        \left\{ \frac{e+1}{p} \mathcal{B} \sqrt{\frac{-\mathcal{D}}{\mathcal{C}}}{F} \left[i \text{ arcsinh} \left(\sqrt{\frac{-\mathcal{B}}{\mathcal{A}}} \tan\frac{\xi}{2}\right), \frac{\mathcal{A}\mathcal{D}}{\mathcal{B}\mathcal{C}}\right]   \right.\\
        & \left.  +\left(\frac{8M}{p}-2\right) \sqrt{\frac{-\mathcal{B}}{\mathcal{A}}} {\Pi } \left[ \left(\frac{e-1}{e+1}\right)^2, i \text{ arcsinh}\left(\sqrt{\frac{-\mathcal{D}}{\mathcal{C}}} \tan\frac{\xi}{2}\right), \frac{\mathcal{B}\mathcal{C}}{\mathcal{A}\mathcal{D}}\right] \right.  \\
        &\left.   + \left(e^2 -1 \right) \frac{4M}{p} {\Pi }\left[\frac{-\mathcal{C}}{\mathcal{D}}, i \text{ arcsinh}\left(\sqrt{\frac{-\mathcal{D}}{\mathcal{C}}} \tan \frac{\xi}{2}\right), \frac{\mathcal{B}\mathcal{C}}{\mathcal{A}\mathcal{D}}\right]
        \right\}  +   \frac{2  {F} \left(\frac{\xi}{2}, \frac{4 e M}{\mathcal{A}}\right)\sqrt{p}}{\sqrt{-\mathcal{A}}},\label{budingjifen}
        \end{aligned}
    \end{equation}
\end{widetext}
where we have introduced notations
\begin{equation}
\begin{aligned}
        & \mathcal{A} \equiv 2 e M+6 M-p, \quad \mathcal{B} \equiv 2 e M-6 M+p, \\
        & \mathcal{C} \equiv 2 e M+2 M-p, \quad \mathcal{D} \equiv 2 e M-2 M+p.
\end{aligned}
\end{equation}
${F}$ and ${\Pi }$ denote the incomplete elliptic integral of the first kind and the third kind, respectively, i.e.
\begin{equation}
\begin{aligned}
        &F(\varphi ,k)=\int _{0}^{\varphi }\frac{dx}{\sqrt{1-k^{2}\sin^{2} x}}, \\
        &\Pi (n;\varphi ,k)=\int _{0}^{\varphi }\frac{dx}{\left( 1-n\sin^{2} x\right)\sqrt{1-k^{2}\sin^{2} x}},
\end{aligned}
\end{equation}
where $k$ is the elliptic modulus satisfying $0< k^{2} < 1$, $n$ is the elliptic characteristic. According to Eqs.~\eqref{deltaAB} and \eqref{schdelta}, for two arbitrary points identified by $\xi_A$ and $\xi_B$ (assuming $\xi_B>\xi_A$), the deflection angle along the trajectory from $A$ to $B$ is expressed as the increment of $\Omega$
\begin{equation}
    \delta_{BA}  =\Omega\left(\xi_B\right)-\Omega\left(\xi_A\right),
    \label{jihepianzhuanjiao}
\end{equation} 
which is an analytical result without any approximation.

\subsection{Discussion about the deflection angle}
Consider the expression of the deflection angle Eq.~\eqref{deltapsiphi}, $\delta^\Psi_{AB}$ describes the increment of the angle between the velocity direction and the radial direction, which is related to the first term of Eq.~\eqref{budingjifen}. $\delta_{BA}^\phi$ describes the increment of the azimuthal coordinate which is related to the second term of Eq.~\eqref{budingjifen}. 

\begin{figure}[!ht]
	\includegraphics[width=0.9\columnwidth]{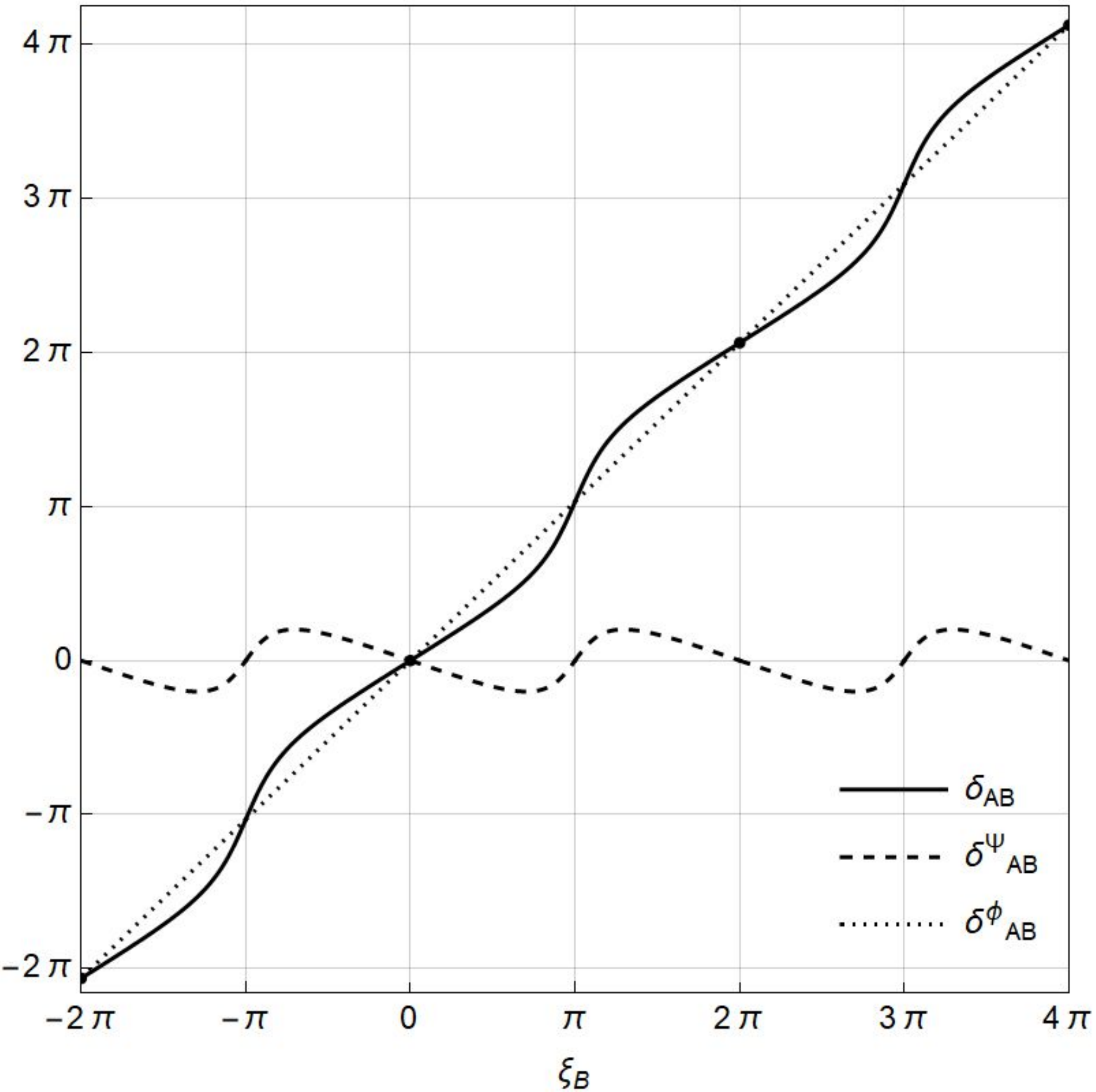}
	\caption{ $\delta_{BA}$, $\delta^{\Psi}_{AB}$ and $\delta^{\phi}_{AB}$ against $\xi_B$ with $\xi_A=0$, $e=0.6$, and $p=10^2M$.}
	\label{fig-7}
\end{figure}

The $\delta_{BA}$, $\delta_{BA}^\Psi$ and $\delta_{BA}^\phi$ against $\xi_B$ are plotted in Fig.~\ref{fig-7}, where we assume $\xi_A=0$, i.e. point $A$ is a pericenter.
The first term of Eq.~\eqref{budingjifen} can be seen as a composite function, $\xi$ only exists in the inner function $\tan\left( \xi/2 \right)$ whose period is $2\pi$ and the other outer functions is monotonic. Thus we can see that $\delta_{BA}^\Psi$ oscillates with a period of $2\pi$.
When the particle moves from a pericenter to the next adjacent pericenter, $\delta_{BA}^\Psi$ decreases slowly from zero to a local minimum, then it increases quickly near the apocenter, finally it decreases slowly back to zero after a local maximum point.
For all pericenters and apocenters, the velocity direction is perpendicular to the radial direction, namely, $\Psi\left(\xi=0,\pm\pi,\pm2\pi,\cdots\right)=\pi/2$, thus $\delta^\Psi_{AB}\left(\xi_B=0,\pm\pi,\pm2\pi,\cdots\right)=0$.
$\delta^\Psi_{AB}$ is more sensitive to $\xi_B$ near apocenters than pericenters, this is consistent with our intuition that for an elliptic orbit, the farther the particle is from the focal point, the quicker the angle between velocity direction and radial direction changes with the same "rotation amplitude".
$\delta_{BA}^\phi$ monotonically increases as $\xi_B$ increases. When the particle moves from a pericenter to the next pericenter, for instance, $\xi_B$ varies from $0$ to $2\pi$, $\delta_{BA}^\phi$ is a little bigger than $2\pi$, the difference between them is the pericenter advance angle.

$\delta_{BA}$ is dominated by $\delta^\phi_{AB}$, and $\delta^\Psi_{AB}$ contributes a periodic perturbation to $\delta_{BA}$. Specifically, $\delta^\Psi_{AB}$ have a negative effect on $\delta_{BA}$ in the first half cycle, while positive effect in the second half.
We emphasize again that $\phi$ represents a monotonically increasing azimuthal coordinate along the trajectory, thus $\delta^\phi_{AB}$ may be greater than $2\pi$, so does $\delta_{BA}$ . For the zoom-whirl orbits with more than one winding within a radial cycle, the behavior of $\delta_{BA}$, $\delta_{BA}^\Psi$ and $\delta_{BA}^\phi$ will be a little more complicated.

The analysis about the dependence of $\delta_{BA}$ on the orbit parameters $e$ and $p$ is shown in the Appendix.

\section{conclusion}
Based on the Jacobi metric, we project the bound orbit of massive particles moving in the equatorial plane of four-dimensional SSS spacetimes into a two-dimensional Jacobi space. (a) Since the bound trajectory of massive particles is periodic and will overlap with itself azimuthally, we divide it into multiple segments such that each segment does not overlap with itself azimuthally. Then those segments are treated as unbound trajectories, and the finite-distance deflection angle for unbound trajectories defined by Ishihara $et\ al.$ is extended to bound trajectories. In addition, the discussion of the deflection angle for unbound trajectories from the view of observation is also extended to the bound scenario. (b) For the bound massive particle, there exists a singular region in the Jacobi space due to the energy per unit rest mass being less than $1$. Thus the infinite region constructed for unbound trajectories is useless for the bound scenario. Thanks to the generalized GW method, the auxiliary circle arc can be chosen freely to construct the region that the GBT can be applied to, and a very simple calculation formula is obtained for two arbitrary points on bound trajectories. We successfully extend the GW method to the bound trajectory of massive particles. Finally, to show the application of the extended GW method, we compute the deflection angle along the bound trajectory between two arbitrary points for massive particles in Schwarzschild spacetime, and obtain an analytical result.

\section*{Appendix: The dependence of $\delta_{BA}$ on $e$ and $p$}
\label{app1}
Let's see how the parameters $e$ and $p$ affect the deflection angle $\delta_{BA}$.

\begin{figure}[!ht]
	\includegraphics[width=0.9\columnwidth]{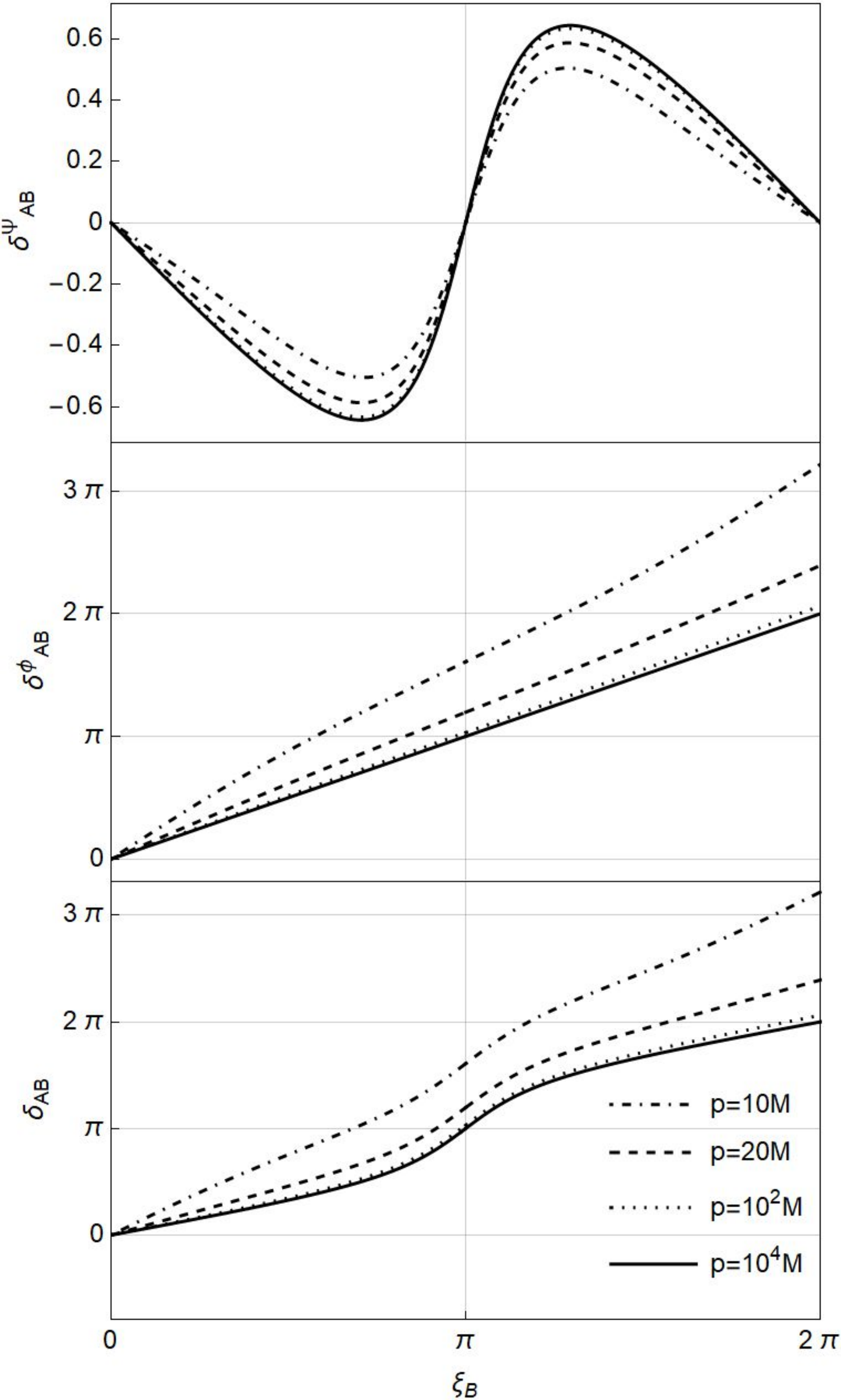}
	\caption{$\delta_{BA}$, $\delta^{\Psi}_{AB}$ and $\delta^{\phi}_{AB}$ against $\xi_B$ with $\xi_A=0$, $e=0.6$, and $p=10M,20M,10^2M,10^4M$. }
	\label{fig-8}
\end{figure}

\begin{figure}[!ht]
	\includegraphics[width=0.9\columnwidth]{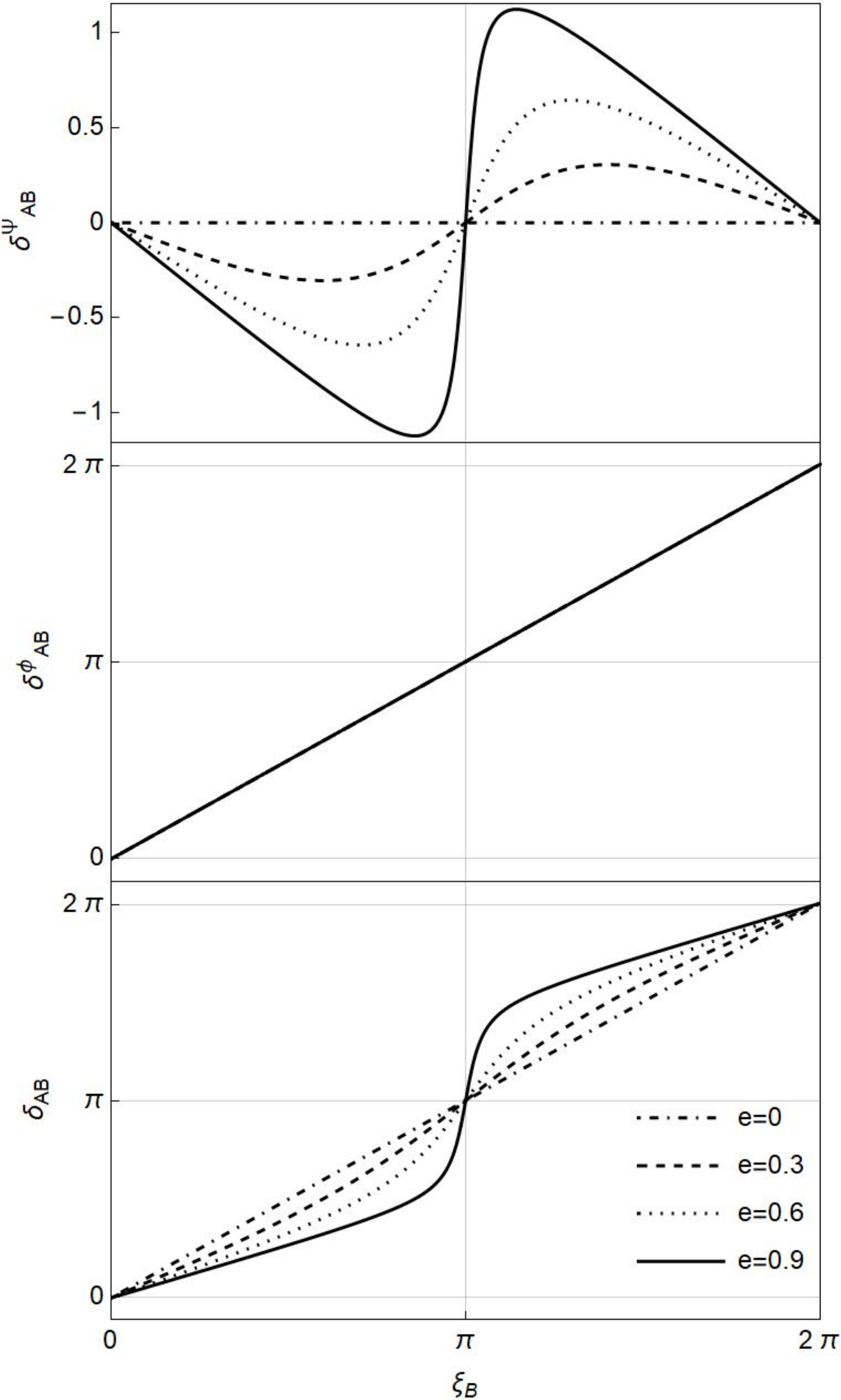}
	\caption{$\delta_{BA}$, $\delta^{\Psi}_{AB}$ and $\delta^{\phi}_{AB}$ against $\xi_B$ with $\xi_A=0$, $e=0,0.3,0.6,0.9$, and $p=10^2M$. }
	\label{fig-9}
\end{figure}

The behavior of $\delta_{BA}$, $\delta^{\Psi}_{AB}$ and $\delta^{\phi}_{AB}$ in a radial cycle is shown in Fig.~\ref{fig-8} for which we fix $e=0.6$ and assume $\xi_A=0$.
The bigger the semi-latus rectum $p$, the bigger the absolute value of $\delta^\Psi_{AB}$ and the smaller the $\delta^\phi_{AB}$. But when $p$ is bigger than $100M$, $\delta^\Psi_{AB}$ and $\delta^\phi_{AB}$ will saturate. $\delta_{BA}$ is dominated by $\delta^\phi_{AB}$.

Fig.~\ref{fig-9} shows the $\delta_{BA}$, $\delta^{\Psi}_{AB}$ and $\delta^{\phi}_{AB}$ against $\xi_B$. We fix $p=100M$ and assume $\xi_A=0$. As we can see, the bigger the eccentricity $e$, the bigger the absolute value of $\delta^\Psi_{AB}$. When $e$ increases, the local minimum point and local maximum point move towards the apocenter at the same time, and consequently, $\delta^\Psi_{AB}$ varies more quickly near the apocenters. When $e=0$, the trajectory reduces to a circle whose velocity direction is perpendicular to the radial direction at all points, accordingly $\delta^\Psi_{AB}$ is identically equal to zero. 
Additionally, $e$ has little effects on $\delta^\phi_{AB}$. Thus, the relation between $\delta_{BA}$ and $p$ is similar to that between $\delta^\Psi_{AB}$ and $p$, although $\delta^\phi_{AB}$ dominates the $\delta_{BA}$.

\bibliography{refs}

\end{document}